# Design Insights for Industrial CO$_2$ Capture, Transport, and Storage Systems

*Tubagus Aryandi Gunawan*[1*], *Lilianna Gittoes*[1], *Cecelia Isaac*[1], *Chris Greig*[1], *Eric Larson*[1]

[1]Andlinger Center for Energy and the Environment, Princeton University

*Corresponding author e-mail: arya.gunawan@princeton.edu

## Abstract

We present design methods and insights for CO$_2$ capture, transport, and storage systems for clusters of industrial facilities, with a case-study focus on the state of Louisiana. Our analytical framework includes: (1) evaluating the scale and concentration of capturable CO$_2$ emissions at individual facilities for the purpose of estimating the cost of CO$_2$ capture retrofits, (2) a screening method to identify potential CO$_2$ storage sites and estimate their storage capacities, injectivities, and costs; and (3) an approach for cost-minimized design of pipeline infrastructure connecting CO$_2$ capture plants with storage sites that considers land use patterns, existing rights-of-way, demographics, and a variety of social and environmental justice factors. In applying our framework to Louisiana, we estimate up to 50 million tCO$_2$/y of industrial emissions (out of today's total emissions of 130 MtCO$_2$/y) can be captured at under \$100/tCO$_2$ and up to 100 MtCO$_2$/y at under \$120/tCO$_2$. We identified 98 potential storage sites with estimated aggregate total injectivity between 330 and 730 MtCO$_2$/yr and storage costs ranging from \$8 to \$17/tCO$_2$. We find dramatic reductions in the aggregate pipeline length and CO$_2$ transport cost per tonne when groups of capture plants share pipeline infrastructure rather than build dedicated single-user pipelines. Smaller facilities (emitting < 1 MtCO$_2$/y), which account for ¼ of Louisiana's industrial emissions, see the largest transport cost benefits from sharing of infrastructure. Pipeline routes designed to avoid disadvantaged communities (social and environmental justice) so as not to reinforce historical practices of disenfranchisement involve only modestly higher pipeline lengths and costs. The economics of a phased buildout of an optimal CO$_2$ capture, transport, and storage system design for Southeast Louisiana, which accounts for two-thirds of the state's industrial emissions today, are quantified, including impacts of the 45Q tax credit available under the Inflation Reduction Act of 2022.

## Synopsis

Few methodologies exist for comprehensively optimizing the design and siting of regional industrial CO$_2$ capture, transport, and storage hubs. We develop such a methodology and apply it to Louisiana (USA) to provide new hub-design insights.

# Contents

## 1. Introduction

Prominent modeling studies suggest that deployment of capture, transport and underground storage of $CO_2$ originating from industrial facilities will be essential if decarbonization aspirations embodied in the Paris Agreement are to be achieved [1], [2]. One global study projects the need for capture and storage of more than 5 billion tonnes of $CO_2$ per year by 2050, with 70% of this from industrial facilities and power plants [3]. A comprehensive modeling study for how the United States (U.S.) could achieve net-zero greenhouse gas emissions by 2050 [4] includes several pathways that entail capturing, transporting and storing between 0.9 and 1.7 $GtCO_2/y$ by 2050. A plausible avenue for the development of $CO_2$ capture, transport, and storage (CCTS) in the U.S. and other regions is the establishment of regional CCTS hubs [5], and movement in this direction is evident from a variety of private sector efforts [6], governmental plans [7]–[9] and other published studies (see Supplemental Information, SI.B10).

The work reported here 1) develops a methodology and tools for preliminary design of U.S. CCTS hubs and 2) applies these to a case study region to develop insights that can help facilitate the initial design of hubs in this and other regions. The methodology, summarized in Figure 1, involves combining diverse data with purpose-built and pre-existing (open-source) modeling tools in four steps. First, data on industrial $CO_2$ emitting facilities are gathered to estimate the magnitude and concentration of $CO_2$ emissions at each facility. These are inputs needed for the second step, estimating the capital and operating costs to capture $CO_2$ at each facility. The third step is a screening to identify potential underground $CO_2$ storage sites in the region of interest and then characterizing these sites in terms of storage capacities, injectivities, and costs. The final step involves designing alternative cost-minimized $CO_2$ pipeline layouts for connecting a given set of capture facilities and storage sites, taking into consideration local land use and other siting constraints.

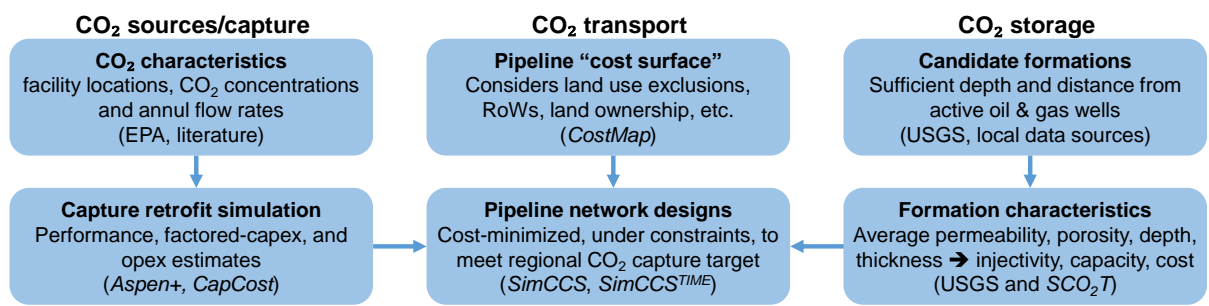

Figure 1. Summary of methodology and associated main data sources and analytical tools used in this study.

*Related work*

Our work adds to a growing literature relating to the design of CCTS hubs and is especially novel in three key respects.

First, we develop a comprehensive methodology and tools for designing cost-minimized hubs for industrial $CO_2$ capture retrofit, pipeline transport, and saline-aquifer storage at high spatial resolution considering all industrial emission sources across a significant (~80,000 km$^2$) geographic area. Prior screening studies have suggested prospectively interesting regions for locating CCTS hubs in the U.S. [10], [11], but did not delve into higher-resolution design of hubs. Most prior studies investigating hubs have limited their focus to a subset of industrial sectors and/or to relatively large individual $CO_2$ emitters, leaving out a wide variety of smaller emitters (< 1 million $tCO_2$/year) that nevertheless collectively might contribute substantially to a region's industrial emissions. Sun et al [12] and Johnsson et al [13] investigated hubs for Spain and Sweden, respectively, but excluded smaller emitters from their analyses. Snyder et al. [14] focus on a limited set of petrochemical industries in the U.S. Edwards et al. [15] analyzed capture of $CO_2$ specifically at corn-ethanol facilities. Pilorgé et al. [16] focused on only a selection of process industries located near existing $CO_2$ pipelines in the U.S. Schmelz et al. [17] and Tao et al. [18] focus exclusively on power plants in the U.S. Fan et al. [19] and Li et al. [20] focus on the power sector in China. Other studies have proposed CCTS hub designs, but without specific optimization of pipeline networks. These include case studies in Louisiana, U.S. [14], Guangdong, China [21], Switzerland [22], Sweden [23], the Iberian Peninsula [12], [24], India [25] and Southeast Asia [26], [27]. And additional studies include optimization of $CO_2$ pipeline routes, but with a narrow focus on specific small industrial zones, including in eastern Ohio [28], western Rotterdam [29], [30] and southeastern Louisiana [31].

Second, our analytical tools for designing integrated CCTS systems include a screening methodology for systematically identifying, through evaluation of both above ground and subsurface features, prospective sites for $CO_2$ injection into saline aquifers and associated estimates of storage capacities, injectivities, and costs for those sites. The large potential storage capacity of saline aquifers will be needed if billion tonne levels of $CO_2$ storage are to be reached globally [32]. A majority of prior studies that have investigated storage of captured $CO_2$ have focused on storage via Enhanced Oil Recovery (EOR), e.g., [17], [33], [34] or depleted oil and gas reservoirs, e.g., [35]. Notably, other authors [36], [37] have suggested that $CO_2$ injection for storage should specifically avoid the proximity of oil and gas exploration and production wells. Our methodology enables avoiding such areas.

Third, our design methodology enables accounting for additional key challenges with siting of $CO_2$ pipelines that prior assessments have not considered. In particular, our approach considers *(i)* private land ownership patterns and land values, which can impact pipeline permitting times and construction and operating costs, and *(ii)* social and environmental justice (SEJ) concerns. Meckel et al. [38] point out the importance of considering permitting processes and requirements in the design of CCTS hubs, because these can contribute to undesirably long lead times for project development. A review by Rubiano et al. [39] notes that SEJ considerations are mostly not integrated in energy modeling frameworks. Süsser et al [40] argue that modeling without considering SEJ factors risks generating overly optimistic and potentially misleading results, and Jenkins et al. [41] and Jenkins et al. [42] indicate that considering them in siting of energy infrastructure will be essential for securing and sustaining social license for the decarbonization transition. Our pipeline routing approach builds on the work of Hoover et al. [43], which considered a foundational set of spatial and cost factors, but not land ownership and value, nor SEJ considerations.

Our methodology is broadly suitable for application to any region of interest. For a case study, we have selected the U.S. state of Louisiana, because several factors point to that state being a potential early mover in the deployment of industrial CCTS hubs: annual statewide $CO_2$ emissions are disproportionately from industry (Figure 2); subsurface geologies are among the most favorable anywhere in the U.S. for $CO_2$ storage [44]; the state has polices supportive of CCTS [45]; and Louisiana's incumbent oil and gas industry has skills and knowledge that are needed for safely building and operating $CO_2$ transport and storage (T&S) systems. Since 2020 here have been at least 14 public announcements of planned $CO_2$ capture and storage projects in Louisiana – see Supplemental Information (SI.A.3).

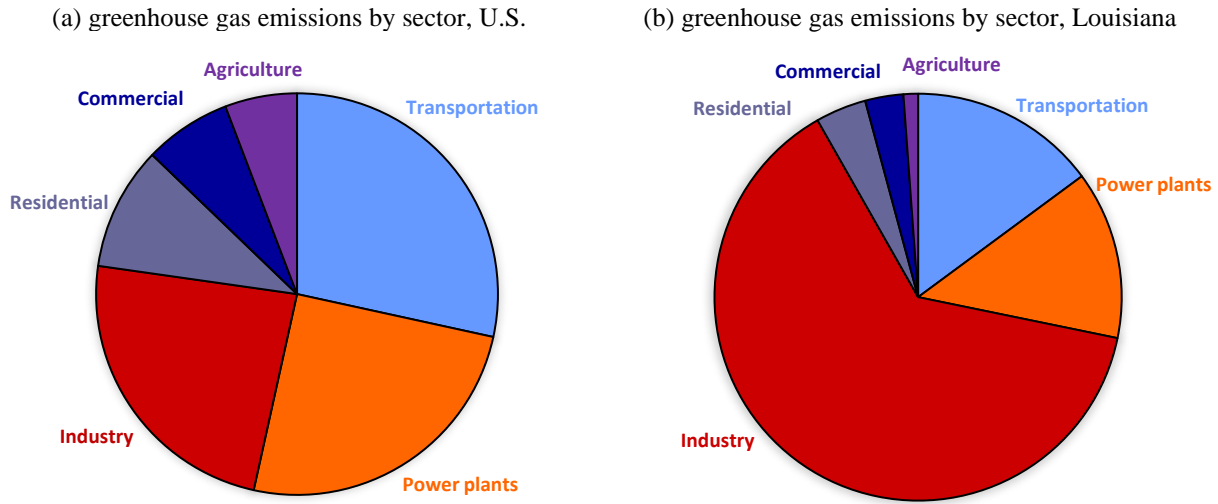

Figure 2. Sector-wise $CO_2$ emissions, 2019: (a) U.S. (total 6,592 million tonnes CO2 equivalent), (b) Louisiana (total 232 million tonnes CO2 equivalent) [46]. Land-use, land-use change, and forestry emissions excluded.

Section 2 presents an inventory of industrial $CO_2$ emission sources and associated capture cost estimates. Section 3 describes our $CO_2$ storage sites screening methodology and results. Section 4 describes our methodology for integrated design of CCTS hubs and presents a series of analyses providing design insights. Finally, Section 5 summarizes our work and suggests important directions for future work.

## 2. Industrial $CO_2$ Emissions and Capture Cost Estimates for Louisiana

An essential input for optimizing the design (layout, sizing, costing) of industrial CCTS hubs is an understanding of the emissions to be captured, including geographic locations of potential capture plants, the rates at which $CO_2$ might be captured at these locations, and the concentrations of $CO_2$ in the gas streams to be targeted for capture. We focus here on inventorying and characterizing emissions from existing industrial facilities in Louisiana. Some of these facilities may be retired in the next few years, but we assume that most of them have long-enough operating lives remaining that $CO_2$ capture retrofits could be considered in the coming decades. Our analysis focuses on Louisiana's emission sources in 2019, the most recent year prior to the Covid-19 pandemic for which complete data are available. In essence, we are assessing CCTS hubs that would be designed for capture of current-day industrial emissions in Louisiana. We do not consider how industrial growth (or shrinkage) in Louisiana might impact industrial emissions, though this is an important topic for future work.

### 2.1. Inventory of Industrial $CO_2$ Emitting Facilities

In 2019 one hundred ninety industrial facilities in Louisiana each emitted more than 25 thousand metric tonnes of $CO_2$, and collectively these facilities emitted 130 million metric tonnes of $CO_2$ (Mt$CO_2$) [47]. Nearly two thirds of these emissions were in the southeast region of the state (Figure 3a), and the largest emitting subsectors in aggregate were natural gas power plants, refineries, and petrochemical plants (Figure 3b).

The largest individual emitters are found in the ammonia, power generation, pulp and paper, and refining sectors, while natural gas and minerals processing facilities are relatively small emitters of $CO_2$ (Table 1). The five largest emitters alone account for 22% of emissions: CF Industries Donaldsonville ammonia manufacturing plant (8.0 Mt$CO_2$/y), ExxonMobil's Baton Rouge refinery (6.3 Mt$CO_2$/y), the Brame Energy Center power plant in Rapides Parish burning coal and natural gas (5.4 Mt$CO_2$/y), Citgo's Lake Charles refinery (4.7 Mt$CO_2$/y), and the Ninemile Point natural gas power plant in Jefferson Parish (4.6 Mt$CO_2$/y).

Three-quarters of total industrial emissions are accounted for by 42 facilities emitting more than 1 Mt$CO_2$/y each. Medium-sized emitters (0.1 to 1 Mt$CO_2$/y each) number 76 and account for an additional 22% of emissions, and the remaining 72 small emitters (< 0.1 MtCO$_2$/y each) account for the remaining 3% of emissions (Figure 4).

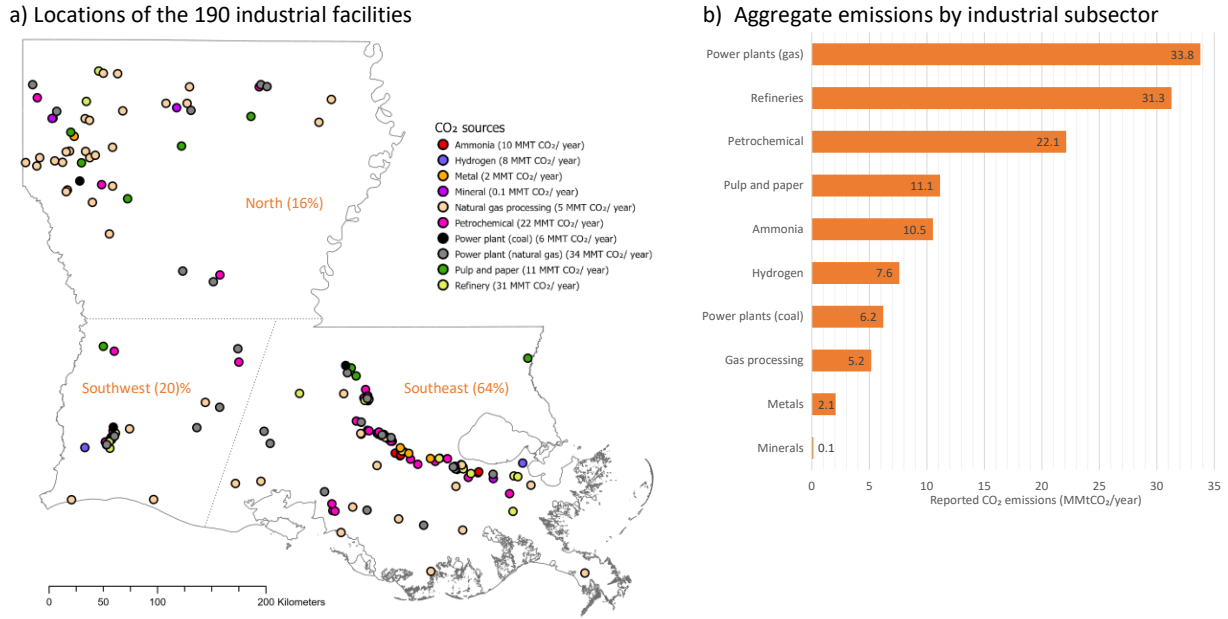

Figure 3. (a) Locations of the 190 industrial facilities that collectively emitted 130 MtCO$_2$ in 2019, with regional percentages as indicated, and (b) corresponding aggregate emissions of industrial subsectors. Data are from [47], excluding three sectors (waste, petroleum and natural gas systems other than natural gas processing, and 'other') and any non-CO$_2$ greenhouse gas emission.

Table 1. Number of individual facilities within industrial subsectors and the range of CO$_2$ emissions (Mt/y) from individual facilities in 2019 [47].*

|  | Number | Smallest | Largest | Average | Total |
|---|---|---|---|---|---|
| **Ammonia** | 4 | 0.531 | 8.016 | 2.630 | 10.52 |
| **Coal power plants** | 3 | 1.149 | 2.910 | 2.069 | 6.207 |
| **Refineries** | 17 | 0.006 | 6.294 | 1.840 | 31.284 |
| **Pulp & paper mills** | 9 | 0.105 | 2.189 | 1.239 | 11.148 |
| **Natural gas power plants** | 28 | 0.011 | 5.389 | 1.207 | 33.796 |
| **Metals** | 4 | 0.031 | 0.978 | 0.515 | 2.062 |
| **Hydrogen** | 15 | 0.016 | 1.072 | 0.507 | 7.603 |
| **Petrochemical** | 56 | 0.018 | 2.903 | 0.395 | 22.12 |
| **Natural gas processing** | 51 | 0.001 | 0.367 | 0.101 | 5.162 |
| **Minerals** | 3 | 0.026 | 0.065 | 0.044 | 0.132 |
| **Total** | 190 |  |  |  | 130.036 |
| * Any facility emitting more than 0.025 MtCO$_2$/y is required to report its emissions to the EPA, and some facilities emitting less than this are also required to report. | | | | | |

**2.2. Characterization of Industrial CO₂ Emission Streams**

In addition to inventorying geographic locations and scales of $CO_2$ emissions, understanding the composition of gas streams from which $CO_2$ can be removed at industrial sites is important, since for a given flow rate of a $CO_2$-laden gas stream, the cost per tonne of $CO_2$ removed generally will be lower for higher $CO_2$ concentrations. Literature sources were consulted to estimate $CO_2$ concentrations for each type of industrial process found in Louisiana, and these are enumerated and documented in SI.B9. For processes that are well-documented, e.g., fossil fuel power plants, ammonia and hydrogen production, and natural gas processing, reliable estimates can be made relatively easily. For not-well-documented processes, such as many chemical manufacturing processes ("Other" in SI.B9), literature-based estimates are not possible, and a simplifying assumption was made, for screening purposes, that emissions would be comparable to those from a natural gas fired-boiler providing process heat.

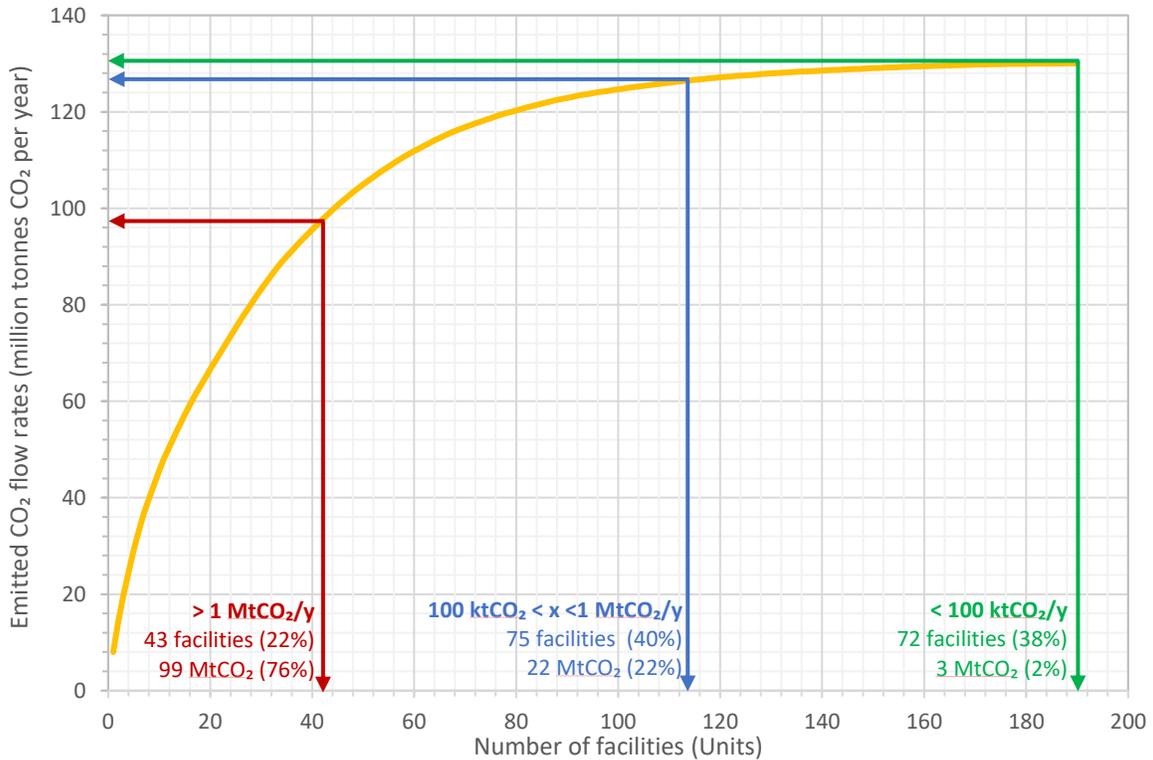

Figure 4. Cumulative $CO_2$ emissions in 2019 from Louisiana's 190 industrial emitters, rank-ordered (left-to-right) from largest to smallest individual emitter.

The estimates in SI.B9 when combined with the facility-level $CO_2$ emissions reported by the U.S. Environmental Protection Agency (EPA) [47] for each of Louisiana's 190 industrial facilities, results in Figure 5, several of which are noteworthy. Ammonia production and natural gas processing facilities both emit $CO_2$ streams containing high purity $CO_2$. These high concentrations are intrinsic characteristics of these processes. Ammonia production involves the intermediate production of hydrogen ($H_2$) by steam methane reforming to produce CO and $H_2$, followed by water gas shift reactions that convert this gas mixture to primarily $H_2$ and $CO_2$, and finally separation of the $H_2$ from the $CO_2$. The resulting $CO_2$ stream requires only dehydration and compression to make it ready for pipeline transport to a storage site. Natural gas processing involves separating $CO_2$ found in raw natural gas in order that the composition of the resulting gas injected into the gas transmission network meets regulatory standards. Much like the case with ammonia, the separated $CO_2$ only requires dehydration and compression to prepare it for pipeline transport.

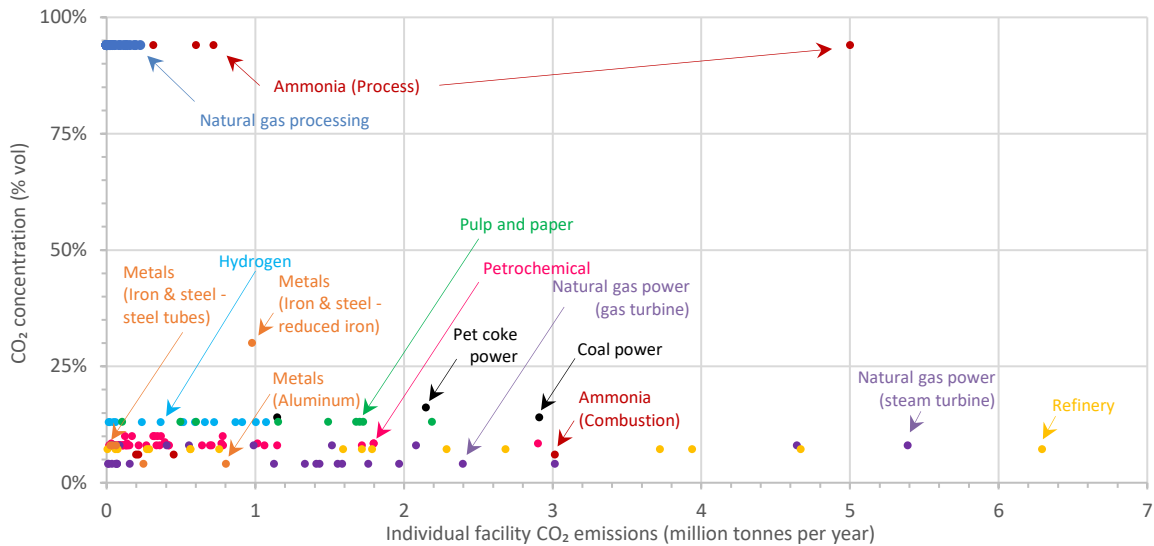

Figure 5. Estimated concentration of $CO_2$ in emission streams at each of Louisiana's 190 industrial emitting facilities plotted against the reported emissions of $CO_2$ from each facility in 2019. $CO_2$ emissions of natural gas processing is only for process emissions.

Except for the high concentration of $CO_2$ in process streams at ammonia and natural gas processing plants, concentrations of other industrial $CO_2$ streams in Louisiana are estimated to be no more than 25%, and most are below 15%, including a secondary emission stream from ammonia plants associated with combustion of fuel to heat the reformer (~5% $CO_2$ concentration). For all such streams, $CO_2$ must first be removed through an added separation process, followed by dehydration and compression before it is ready for pipeline transport to storage sites.

The 3rd largest individual $CO_2$ emission stream in Louisiana today is the process $CO_2$ stream at the CF Donaldsonville ammonia plant (5 million $tCO_2$/y). The high $CO_2$ concentration and large mass flow of this stream make it likely to have the lowest capture cost among any industrial emitter in the state. Other large emitters (> 1 million $tCO_2$/y), which can benefit from scale economies in capture costs, include some natural gas power plants, refineries, coal and petcoke power plants, some petrochemical operations, and pulp and paper mills (Figure 5). In the case of refineries and petrochemical plants, the emissions indicated in Figure 5 represent the sum of a multitude of smaller individual streams at a site, which would add further complication to capturing $CO_2$. Pulp and paper mills are of potential interest due to relatively high $CO_2$ concentrations (among low concentration emitting facilities) and the fact that an estimated 78% of emissions (on average for Louisiana mills) are biogenic emissions (see SI.B1), which holds out the possibility that negative emissions might be achieved with these facilities via CCTS.

## 2.3. Scoping-level cost estimates for $CO_2$ capture at individual facilities

Technologies for $CO_2$ removal from gas streams using amine solvents (e.g., MEA or MDEA) are commercially established [48]. Amine solvents are suitable for $CO_2$ capture at most, if not all, of Louisiana's existing $CO_2$ emitting industrial plants. Gunawan, et al [31] developed Equation 1 for scoping-level estimates of the total levelized cost per tonne of $CO_2$ captured via retrofit of amine-based capture, dehydration, and compression to 110 bar pressure:

$$Capture\ cost\ (2020\$\ /\ metric\ tCO_2) = C_0 \times CO_2\ captured\ (10^6\ metric\ tCO_2\ /\ y)^{-b} \qquad \text{Equation 1}$$

where the values of the coefficient, $C_0$, and the exponent, $b$, vary with the concentration of $CO_2$ in the gas stream entering the amine absorber.[a,b] Table 2 shows parameter values for $N^{th}$-of-a-kind plants. (See [31] for cost estimates of first-of-a-kind and early-mover plants.) Annual $CO_2$ captured is calculated using Equation 2, where $CO_2$ emitted is the $CO_2$ reported emissions by EPA [47], percentage of capturable $CO_2$ emissions for each industrial type are our estimates (see SI.B9), and 0.95 is the design capture fraction.

Table 2. Parameter values for Eqn. (1) [31]. "Nth plant" refers to fully commercially-matured-technology cost levels.

| vol% $CO_2$ → | | 5% | 10% | 15% | 94% |
|---|---|---|---|---|---|
| Nth plant | $C_o$ | 123 | 105 | 99 | 27 |
| | $b$ | 0.146 | 0.167 | 0.175 | 0.415 |

---

[a] Eqn. (1) assumes that 95% of $CO_2$ input to the absorber is captured. It also assumes the capture facility operates with a 90% annual capacity factor and assumes an annual capital charge rate of 10.6%. See Gunawan, et al [31] for other details.

[b] All costs in this paper are in 2020$, unless indicated otherwise.

$$CO_2 \text{ captured } (10^6 \text{ metric } tCO_2 / y) = CO_2 \text{ emitted } (10^6 \text{ metric } tCO_2 / y) \\ \times CO_2 \text{ capturable } (\% \text{ of } CO_2 \text{ emitted}) \times 0.95$$

Equation 2

The cost model (Equation 1) does not include any costs for removal of other contaminants (SOx, NOx, PM) that might be required prior to the gas stream entering the $CO_2$ absorber. One analysis for $CO_2$ capture at a cement plant indicates this might add about \$15/t$CO_2$ captured [49]. Another study [50] suggests that these added costs would be readily paid back from a societal perspective by the monetary value of health benefits (per tonne of $CO_2$ captured) from ceasing emission of these local pollutants. The cost model does include costs for all needed balance of plant equipment, including natural gas fired back-pressure steam turbine cogeneration for meeting internal heat and electricity needs of the capture plant. However, the model also assumes $CO_2$ emissions from the cogeneration system are not captured, which makes the effective annual $CO_2$ capture rate 73% of the value used in Equation 1 to calculate capture cost. (The levelized costs estimated using Equation 1/Table 2 would be slightly lower if facilities were designed to also capture cogeneration $CO_2$ emissions: absolute capital costs would be higher, but scale economies for capture equipment would more than offset this [31].)

Results from Equation 1 for each industrial facility in Louisiana are shown in Figure 6. The majority of emissions are capturable at ($N^{th}$ plant) costs between \$80 and \$120/t$CO_2$. Costs are well below \$80/t$CO_2$ for plants with high (94+ %) $CO_2$ concentrations in the target stream (Figure 6, left-most data points), and they are well above \$120/t$CO_2$ for plants with smaller annual capture rates of target streams with low $CO_2$ concentrations (right-most data points).

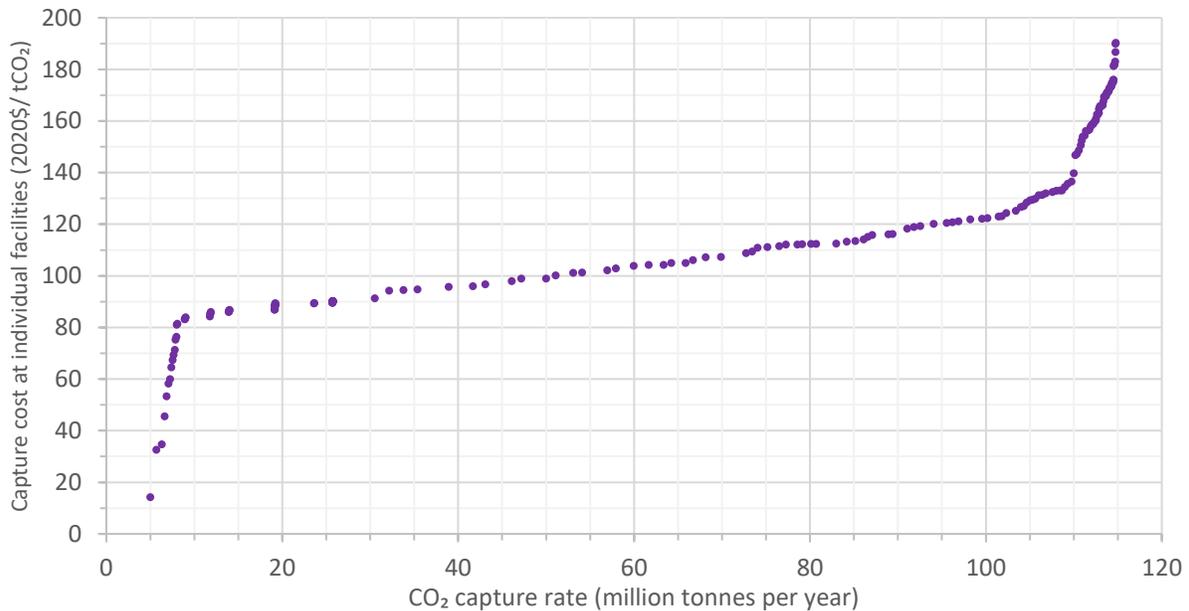

Figure 6. Estimated levelized capture costs for $N^{th}$ plants at individual industrial facilities in Louisiana ordered from lowest to highest. X-axis shows cumulative annual $CO_2$ capture.

## 3. Prospective $CO_2$ Storage Sites and Costs in Louisiana

The Texas/Louisiana Gulf Coast region has extensive subsurface geology prospectively suitable for $CO_2$ storage, both onshore and sub-seabed offshore [36]. We screened Louisiana to identify candidate onshore $CO_2$ injection sites and estimated $CO_2$ injectivities, storage capacities, and storage costs at these sites. This was a high-level screening only, and proper appraisal work would be required to validate any particular site. Once candidate surface sites for injection were identified, physical characteristics of the underlying geological formations, as estimated by the US Geological Survey, provided the basis for estimating prospective $CO_2$ injectivities, storage capacities and costs at each site. We did not screen for offshore $CO_2$ storage sites, but a similar methodology as developed here for onshore sites could be developed for that purpose.

### 3.1. Surface screening

The surface evaluation of onshore areas utilized high-resolution geospatial screening of land cover types (Figure 7a) to identify regions where surface features make it unsuitable for siting $CO_2$ injection wells. All developed areas, roadways, waterways, and water bodies, plus a 20 km buffer region around each of these, were considered exclusion areas for $CO_2$ injection. Additional important features to consider in the case of Louisiana are the 1,863 identified oil and gas fields and nearly one-quarter million associated production and injection wells (Figure 7b). Active oil and gas fields, plus a 5 km buffer around each, were also excluded from being candidate sites for CO₂ injection wells. A final criterion for a surface area to be considered a candidate site for CO₂ injection was that it must be part of a contiguous non-excluded area measuring at least 78.5 km2 (the area of a circle with 5-km radius). This final criterion was included in consideration of the post-injection subsurface lateral migration of the CO₂ plume.

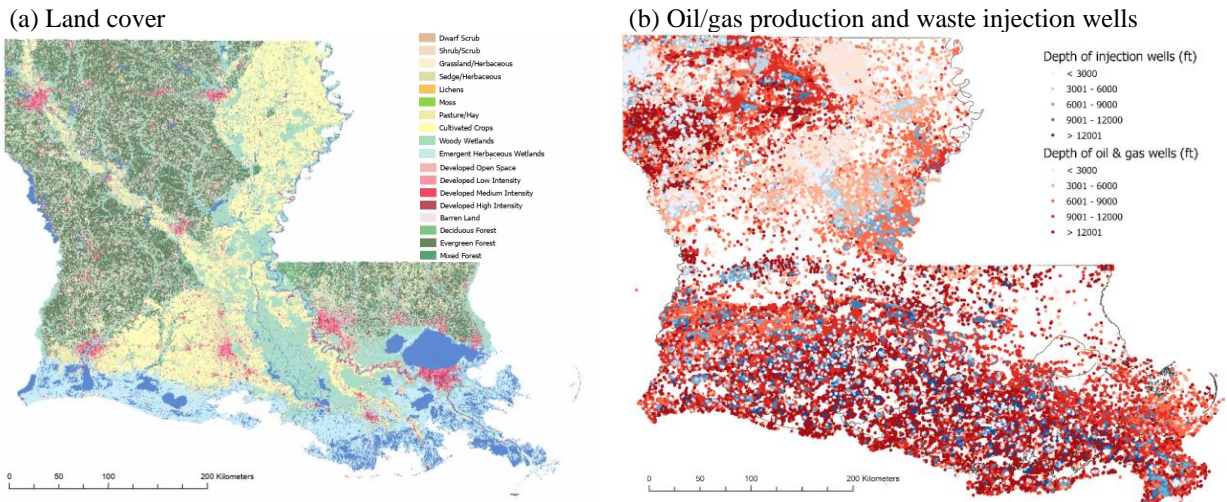

Figure 7. Land surface areas considered in screening for prospective CO₂ storage sites, including (a) land cover type [51] and (b) known oil/gas production and waste injection wells [52]

Our land surface screening analysis identified 87 sites that meet all criteria, as shown in Figure 8. The total number of sites shown is 98, which includes 12 sites not from our screening analysis (red triangles). The latter are sites identified in corporate announcements for CO₂ storage projects – see SI.A3. All 98 sites are considered as candidate storage sites for the purpose of subsequent analyses in this paper.

## 3.2. Subsurface Characterization

The U.S. Geological Survey (USGS) has assessed 27 geological formations in the Gulf Coast region for their suitability for $CO_2$ storage [53]. Of these, 22 underlie Louisiana – see SI.A1. Moving from north to south, the time since the formations were created decreases and correspondingly the prospective suitability for $CO_2$ storage, as judged by porosity and permeability characteristics, increases.

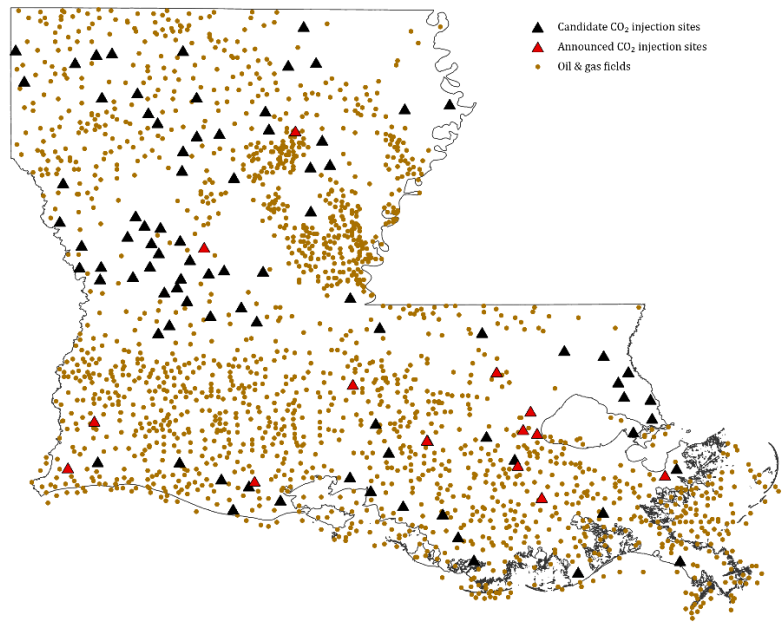

Figure 8. Known oil and gas fields (brown dots) [52], identified candidate $CO_2$ injection sites (black triangles), and $CO_2$ injection sites publicly announced since 2020 (red triangles). See SI.A3 for details of announced sites. Generally, the permeability and porosity underlying candidate sites increase from the north to the south of the state, corresponding to increasingly favorable storage potential. See SI.A1.

Of the 22 formations underlying Louisiana, we considered formations at depths from 3,000 to 13,000 feet to be potential candidates for $CO_2$ storage. Five of the 22 formations are deeper than 13,000 feet, and we eliminated these from consideration. Additionally, the USGS did not report depths for two formations, so those were also excluded, leaving 15 formations considered potentially suitable for $CO_2$ storage. The USGS provides estimates of ranges in depth, thickness, porosity, and permeability for each formation. See SI.A1.

Overlaying results from our land-surface screening in the previous section with the physical extents of the 15 selected formations from the USGS dataset, we associated each candidate storage site on the surface with one or more underlying prospective storage formations. For many candidate storage sites, there are multiple underlying formations that occur at different depths – see SI.A4.

## 3.3. $CO_2$ Storage Costs

The cost of $CO_2$ storage will vary with the $CO_2$ injectivity and storage capacity of a formation, among other variables. Injectivities, capacities, and corresponding estimated storage costs for each candidate storage site were estimated using a customized version of the open-source screening tool, *Sequestration of CO₂ Tool* (*SCO₂T*) [54], [55].

*SCO₂T* first estimates reservoir injectivity (MtCO$_2$/yr) and total storage capacity (MtCO$_2$) for individual storage sites using reduced order models of CO$_2$ plume migration, assuming a cylindrical reservoir having uniform permeability, porosity, temperature and fluid composition, no leakage, and no pressure interference from other injection wells. To account for uncertainties in geological parameter values at each storage site, a range of parameter values were generated using Monte Carlo analyses, following the approach of [55], [56]. Normal distributions for formation depth, thickness, permeability, porosity, and temperature (assuming a geothermal temperature gradient of 32°C/km [55]) were assumed, with mean values as reported by USGS [53] and standard deviations of 10% of the mean (for depth and temperature) and 15% (for the other parameters).[c] The analyses generated 100 randomly sampled values for each parameter. A value from each of the five resulting parameter value distributions was then randomly selected and combined to form 100 sets of parameter values as inputs to the *SCO₂T* tool to calculate injectivity and capacity. The spread in the resulting 100 values of injectivity and capacity is small in most cases (see SI.A9.1 – A9.15), and we assigned the average of the 100 results for each storage site as the injectivity and capacity for that site.[d] The average values for injectivity are shown in Figure 9 for each of the 98 candidate storage sites in Louisiana identified from the surface screening.[e] Injectivities generally increase as one moves from north to south in the state. The collective total estimated injectivity considering all sites is 331 MtCO$_2$/y (Figure 9a). The idea of injecting into multiple, vertically aligned but physically segregated formations at individual sites has been suggested [57], and in this case, the aggregate total injectivity more than doubles to 731 MtCO$_2$/y (Figure 9b). The corresponding aggregate storage capacities are 9,800 MtCO$_2$ and 21,900 MtCO$_2$, respectively. These capacity estimates are significantly less than other published estimates that have not taken into consideration pre-existing oil and gas wells nor surface exclusion areas for siting of injection wells. See SI.A7 for quantitative comparisons.

---

[c] Assuming gamma distributions for parameter values (rather than normal distributions) has negligible impacts on the results. Standard deviation values were selected to help ensure that randomly sampled values generated by the Monte Carlo analysis fall within the validity ranges of the reduced order models embedded in *SCO₂T*.

[d] In some cases, the range of values reported by the USGS exceed the range of validity for the reduced order models embedded in *SCO₂T*. Specifically, maximum thickness and permeability values reported by the USGS for some formations exceed *SCO₂T* validity limits (by a factor of 7 or more). In such cases, we (conservatively) reduced the USGS range to match the *SCO₂T* limits, which reduces injectivity and capacity estimates relative to those with higher assumed permeability and thickness limits.

[e] Two of the candidate storage sites identified in the surface screening work were found to have no underlying storage formations and so we eliminated those from consideration as potential storage sites.

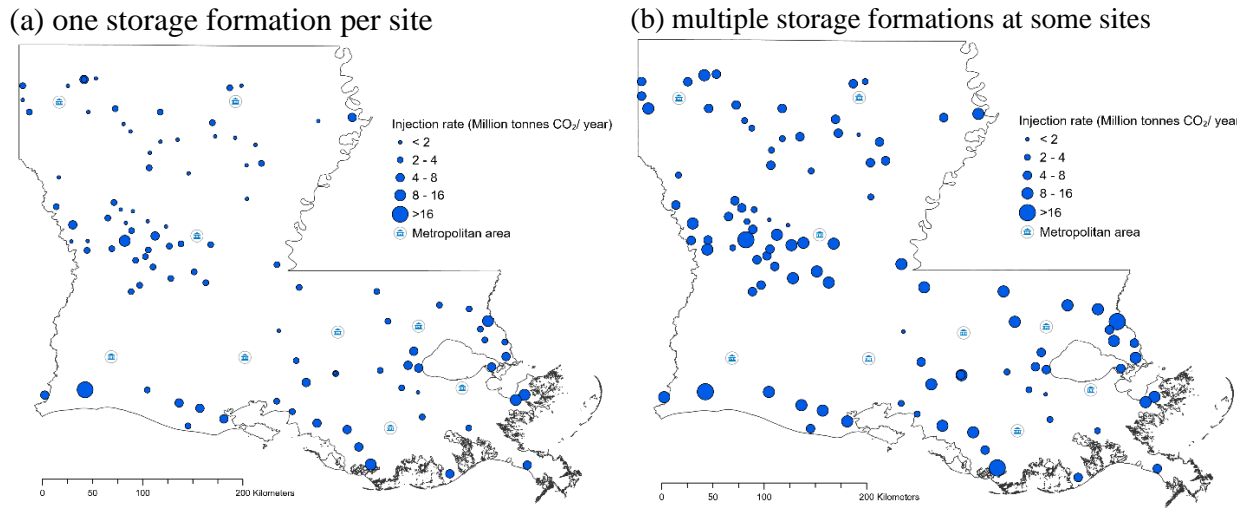

Figure 9. Calculated $CO_2$ injectivities at storage sites, assuming each site accesses (a) only one storage formation and (b) multiple storage formations (at different depths).

*$SCO_2T$* estimates $CO_2$ storage costs for a site, based on the injectivity and capacity values estimated from its embedded reduced order reservoir models. The estimated storage costs include site characterization, well construction, testing, operation, monitoring and post-injection plugging, and site remediation. They do not include any lease payment for use of the subsurface. (See SI.A8 for details of unit-cost assumptions used in *$SCO_2T$*.) A single storage site typically encompasses multiple injection wells to achieve the estimated site injectivity level. Additional input assumptions for the *$SCO_2T$* cost modeling include a maximum injectivity of 1 $MtCO_2/y$ per well, one injection pump per well, one water production well for each injection well, formation pressurization not to exceed 80% of estimated formation fracture pressure, $2/m^3$ water treatment/disposal costs, a 30-year injection period, and a 15% discount rate. The assumed areal extent of a storage reservoir is the contiguous area ($\geq$ 78.5 $km^2$) connected to each storage site identified through the surface screening described in Section 0.

When injection is considered into the single best (lowest-cost) formation at each candidate site, the storage cost ranges from a low estimate of 8.3 $/t$CO_2$ for sites in the Lower Miocene II formation in Southeast Louisiana to a high of 11.7 $/t$CO_2$ for sites in the Norphlet formation in the North of the state. When storage in multiple, vertically aligned formations at individual sites is considered, the average cost at a site ranges from 8.3 $/t$CO_2$ to 16.7 $/t$CO_2$ (Figure 10). In calculations in the next section, we limit consideration to storage only in the single formation at each site. For sites with multiple stacked formations, we select the one with the lowest estimated storage cost.

## 4. Analysis of $CO_2$ Capture, Transport, and Storage Systems

Utilizing the data sets described above, we developed insights to help inform the design of CCTS systems for industrial facilities in Louisiana by exploring alternative pipeline networks that could link $CO_2$ capture points to storage sites.

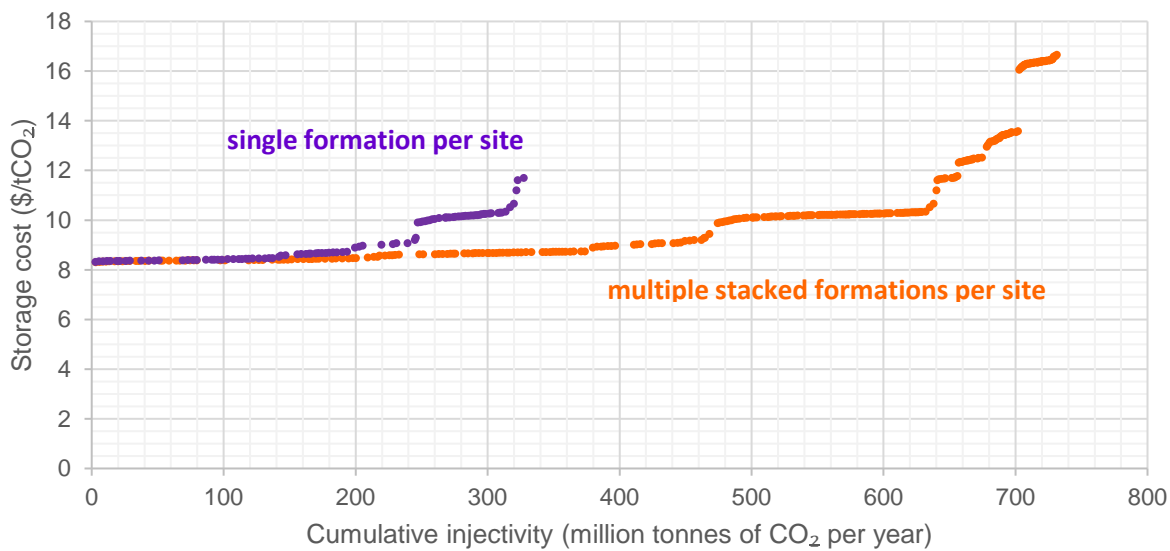

Figure 10. $CO_2$ storage resource cost-supply estimates. See SI.A9 for details.

### 4.1. Approach

We assume $CO_2$ is captured at existing industrial facilities (Figure 3a) at costs estimated in Figure 6, and that $CO_2$ storage occurs at candidate storage sites (Figure 8) with estimated injectivities and costs as in Figure 9a and Figure 10 (single formation injection), respectively. We input these data sets to $SimCCS^{PRO}$, a customized version of an open-source mixed-integer linear program for cost-optimizing capacities and routings of $CO_2$ pipeline networks connecting $CO_2$ sources and sinks, taking into consideration local land use and other pipeline siting constraints [58]. The $SimCCS^{PRO}$ algorithm minimizes the total capture, transport, and storage system cost over a specified operating lifetime for a given set of user-defined $CO_2$ sources and sinks and a target total amount of $CO_2$ capture and storage within a geographic region. A base pipeline cost model that encompasses pipeline materials, construction, right-of-way acquisition, and operating costs under a standard set of reference conditions [59] is embedded in *SimCCS$^{PRO}$* to calculate base pipeline costs as a function of length and flow capacity. Costs are included for pumping stations located every 80 km along a pipeline (to maintain a $CO_2$ pressure between 86 and 150 barg).

Additionally, a cost surface embedded in *SimCCS$^{PRO}$* [43] adjusts the base cost of each segment of a pipeline according to user-assigned weightings of different topographic and demographic features encountered by that segment of pipeline. The native cost surface in *SimCCS$^{PRO}$* delineates (at 60-120 meter resolution) land cover types, federal land areas, existing rights-of-way (natural gas pipelines, electricity transmission, railways, roadways), terrain slopes, and rivers. We added land ownership boundaries and estimated land values to the native cost surface[f] to enable the design of pipelines that favor traversing lower-value lands and that minimize the number of distinct owners that a pipeline developer would need to interact with for land acquisition and permitting. We assigned weights to these and all other features of the cost surface, as detailed in SI.B2.

Finally, for some analyses, we delineated communities historically marginalized, underserved, or overburdened by pollution by adding a social and environmental justice (SEJ) layer to the cost surface. One could assign weightings to SEJ areas that either encourage or discourage pipeline network designs impinging on these areas. On the one hand, $CO_2$ pipeline projects may benefit communities by offering economic opportunities and jobs [60]. On the other hand, some communities may feel such projects are undesirable [61], [62]. Here, to illustrate how consideration can be given to SEJ areas in pipeline network design, we assign weightings that discourage pipeline siting in SEJ areas. We created the SEJ cost surface layer by overlaying a 90m x 90m resolution population dataset map [63] with a dataset of disadvantaged census tracts as identified by the Climate and Economic Justice Screening Tool (CEJST) [64]. Any 90x90 square with population greater than five that overlapped with a disadvantaged census tract was included in our SEJ layer, which allowed rural disadvantaged communities to be included in the layer while leaving out large unpopulated areas within a CEJST census tract. Additionally, a buffer of 182m was included around each 90x90m square to realistically model setbacks, and to prevent the CostMAP software from creating pipeline routes crossing between semi-isolated population squares, which would be counter to the goal of creating distance from disadvantaged communities. Heavy weightings were assigned to the included 90x90m squares to discourage pipelines being routed through these by *SimCCS* – see SI.B2. In essence, our routing algorithm only allows pipelines in disadvantaged areas if there is no physical alternative, for example if a capture plant itself is located in a disadvantaged area. It is worth noting that when $CO_2$ capture is added at an industrial facility, health benefits are likely to accrue in nearby surrounding areas due to concomitant reductions in emissions of local pollutants [50].

---

[f] Land parcel geometries and values were obtained from the parcel data company Regrid (www.Regrid.com) via their academic and nonprofit oriented Data With Purpose program (www.Regrid.com/Purpose).

Using the above approach, we created two SEJ layers. A layer we designate SEJ8 considers census tracts having at least one of any of the eight categories of burdens that defines a census tract as disadvantaged in the CEJST: health, water/wastewater, legacy pollution, climate change, energy, housing, transportation, and workforce development. The second SEJ layer we constructed, SEJ3, considers only census tracts having health, water/wastewater, and/or legacy pollution burdens, the categories potentially most likely to be directly exacerbated by the construction and operation of $CO_2$ pipelines.

## 4.2. Results

Our analyses described in this section provide quantitative techno-economic insights into $CO_2$ capture, transport, and storage system designs that can help inform planning, but the results presented here should be considered illustrative, rather than definitive for several reasons. First, *SimCCS* optimizes pipeline routings with comprehensive knowledge of all possible routes, perfect foresight regarding target capture levels, and instantaneous construction of physical plants and infrastructure. Additionally, the modeling is idealized in that it does not consider many, largely non-technical, factors that can influence network designs [65]. Finally, computational optimization tools, like *SimCCS*, can produce different physical network designs with very similar overall costs.

### 4.2.1. One-user pipelines

Most announced $CO_2$ capture and storage projects in Louisiana involve a capture site connected to a storage site with a dedicated pipeline not shared with any other user. To assess the implications of this strategy becoming the norm for future CCS deployments, we first ran a *SimCCS*[PRO] optimization where $CO_2$ is captured from each of the 190 industrial sources[g] in

---

[g] In this analysis, each of the four ammonia plants in Louisiana is considered as two sources of $CO_2$, one being high purity process $CO_2$ and the other being low purity flue gas $CO_2$.

Louisiana and transported to storage sites in pipelines dedicated to serving only one individual source each. Subject to that constraint *SimCCS$^{PRO}$* identifies the storage site that provides the lowest total transportation and storage (T&S) cost per tCO$_2$ for that source. (In our modeling, multiple pipelines were allowed to terminate at a common storage site if the storage formation at that site has adequate injectivity and capacity.[h]) For this state-wide analysis, the total estimated overnight capital investment in (N$^{th}$ of kind) capture plants was $11.7 billion. Aggregate pipeline and storage capital investment was $1.7 billion and $1.0 billion, respectively. Close examination of a few individual sources and their associated T&S infrastructure (Table 3) offers the following insights:

- Except for the four CO$_2$ sources from natural gas processing, denoted '(ngp)' in the table, unit capture costs decrease with increasing scale of capture, as expected. (The 94%+ CO$_2$ concentration in gas processing capture streams leads to lower capture costs compared with those at facilities that are larger, but where CO$_2$ concentrations are much lower.)
- Capture costs dominate total levelized CCTS costs for all but the very smallest facilities in Table 3. This fact, combined with economies of scale with increasing capture plant capacity, suggests that industrial facilities in close proximity to each other (especially smaller facilities) that pool their CO$_2$ streams and share a capture plant may enjoy capture-cost benefits over capture plants at individual facilities. This idea is explored in detail by Gunawan, *et al*. [31].
- Scale economies in pipeline costs are also evident in Table 3. In the case of the smallest facility in each of the three regions, high unit transport costs associated with small pipeline capacities are exacerbated by the presence of certain geographic or demographic features. For example, the pipelines from the Delhi and Patterson natural gas processing plants, respectively, pass through a highly developed area and across a river, the cost weightings for which drive up pipeline cost significantly (SI.B2).

---

[h] The siting of dedicated pipelines was carried out as follows. First, the largest single CO$_2$ source in Louisiana was connected to a storage site with a dedicated pipeline, with *SimCCS$^{PRO}$* determining the storage site and pipeline routing that minimizes total T&S cost. The available injectivity at the selected storage site was then decremented correspondingly. The closest CO$_2$ source (straight-line distance) to the largest source was then connected to a storage site with its own dedicated pipeline using *SimCCS$^{PRO}$* in a similar manner. Additional sources were then selected for matching with storage sites according to increasing straight-line distance from the largest source, until all sources had been matched with storage. Throughout the matching process when a storage site reached its estimated injectivity limit, that site was no longer available to connect with any additional sources.

- Within a given region of the state (north, southwest, or southeast), levelized storage costs are essentially insensitive to the rate of $CO_2$ storage at a site, but storage costs are consistently higher in the north than in the southwest and southeast regions due to more favorable storage geology in the southern part of the state, as discussed earlier.

Table 3. Capture ($N^{th}$ plant), transport, and storage (CTS) costs for individual $CO_2$ sources using dedicated pipelines (not shared with any other source). Shown here are $SimCCS^{PRO}$ calculated values for facilities with minimum, first quartile, median, third quartile, and maximum annual $CO_2$ capture rates among all facilities in each region. See SI.B1 for details of these, and other facilities, in Louisiana.

| $CO_2$ source | Capture, $MtCO_2$/y | Levelized cost, 2020$/metric tonne $CO_2$ | | | | Pipe km | Trpt. cost, $/$10^3$ t-km |
|---|---|---|---|---|---|---|---|
| | | Capture | Transport | Storage | Total | | |
| *North region (see Figure 3(a) for region definitions)* | | | | | | | |
| Delhi NGL (ngp[a]) | 0.00004 | 90 | 2,056 | 11 | 2,157 | 36 | 57,100 |
| Calumet Lubricants (refinery) | 0.02090 | 165 | 91 | 12 | 268 | 18 | 5,060 |
| Lieberman Power | 0.04800 | 171 | 42 | 13 | 226 | 9 | 4,670 |
| Magnolia Amine (ngp[a]) | 0.18600 | 60 | 17 | 11 | 89 | 89 | 190 |
| Brame Energy Center | 5.11980 | 87 | 1 | 10 | 97 | 21 | 50 |
| *Southwest region* | | | | | | | |
| Targa Downstream (ngp[a]) | 0.00007 | 90 | 2,003 | 9 | 2,102 | 79 | 26,600 |
| Calcasieu power plant | 0.06720 | 182 | 134 | 9 | 325 | 79 | 1,700 |
| Cabot Corp | 0.34860 | 127 | 11 | 9 | 147 | 27 | 410 |
| Coughlin Power Station | 1.47680 | 120 | 5 | 9 | 134 | 31 | 160 |
| Citgo Petroleum Corp | 3.54860 | 96 | 5 | 9 | 109 | 84 | 60 |
| *Southeast region* | | | | | | | |
| Patterson (ngp[a]) | 0.00004 | 90 | 3,567 | 9 | 3,665 | 99 | 36,030 |
| Houma | 0.03500 | 173 | 231 | 8 | 412 | 43 | 5,370 |
| Almatis Burnside | 0.23710 | 156 | 12 | 8 | 177 | 14 | 860 |
| Noranda Alumina | 0.76320 | 132 | 8 | 8 | 149 | 24 | 330 |
| CF Industries Donaldsonville[b] | 5.00090 | 14 | 4 | 8 | 26 | 74 | 55 |

(a) Natural gas processing (94% $CO_2$ concentration in capture target stream).
(b) Ammonia plant. Costs here are associated only with the process $CO_2$ target capture stream (94% $CO_2$ concentration).

### 4.2.2. One-user pipelines vs. shared infrastructure

While capture costs generally dominate total levelized costs, reducing the extent of T&S infrastructure needed for a given level of total annual $CO_2$ storage by sharing infrastructure among multiple facilities nevertheless can result in cost reductions and, at least as importantly, facilitate deployment of T&S systems by reducing siting/permitting challenges and construction times. An illustration of the potential benefits can be seen by comparing two state-wide T&S system designs for the same total amount of $CO_2$ captured and stored. In one design individual sources are constrained to use a dedicated (unshared) pipeline to storage sites (as in Section 4.2.1). In the other design, sources are allowed to share pipelines if it reduces total T&S costs. Figure 11 visualizes the optimal T&S networks with these two strategies when all existing industrial sources in Louisiana are included.[i]

---

[i] When considering shared T&S infrastructure, the design of that infrastructure may vary with the number and size of capture plants sharing the infrastructure. Here we develop insights on T&S infrastructure designs when considering $CO_2$ capture at all industrial facilities in Louisiana.

Figure 11. (a) One-user pipelines and (b) optimized shared pipeline networks. Some southwestern $CO_2$ pipelines are routed off- and then back on-shore to avoid traversing wetlands while following existing oil & gas pipeline rights-of-way. Each ammonia facility is modeled as two separate $CO_2$ sources, as discussed in the text.

(a) single-user pipelines

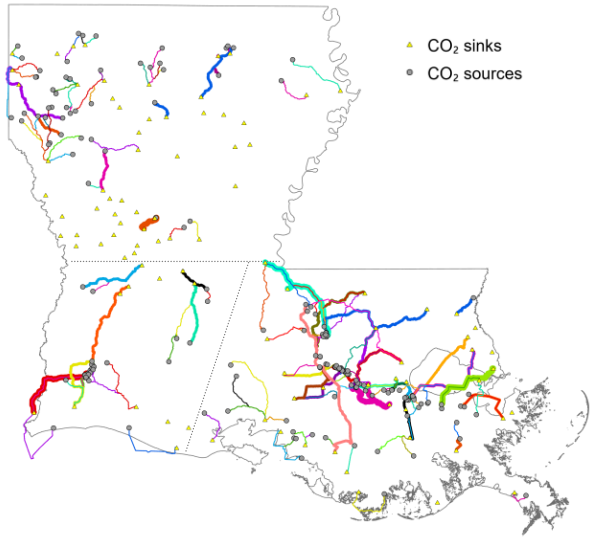

(b) optimized sharing of T&S

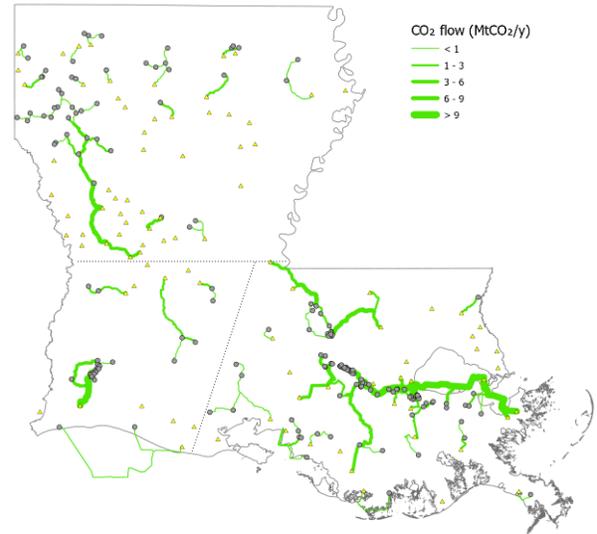

| | |
|---|---|
| Total annual capture (MtCO$_2$/y) | 115 |
| Number of capture units | 194 |
| Length of pipelines (km) | 10,100 |
| Number of storage sites | 63 |
| Average levelized cost ($/tCO$_2$) | |
|   Capture | 102 |
|   Transport | 15 |
|   Storage | 9 |
|   Total | 126 |
| Total annual capture (MtCO$_2$/y) | 115 |
| Number of capture units | 194 |
| Length of pipelines (km) | 2,858 |
| Number of storage sites | 45 |
| Average levelized costs ($/tCO$_2$) | |
|   Capture | 102 |
|   Transport | 6 |
|   Storage | 9 |
|   Total | 116 |

Capture costs are the same in both cases, but when T&S infrastructure is shared, total pipeline length is reduced by more than 75%, the average unit cost for transport is reduced by more than 60%, and the number of unique storage sites is reduced by nearly 30%. Table 4 shows that the benefits of shared T&S infrastructure are most significant in the southwest and southeast, with aggregate pipeline length and average transport cost reduced by about 75% and 65% respectively. The corresponding reductions in the north region, where sources and sinks are more remote from each other and less densely clustered, are 53% and 36%.

Table 4. Comparisons of regional metrics with and without shared T&S infrastructure.

| Region of Louisiana → | North | | Southwest | | Southeast | | All | |
|---|---|---|---|---|---|---|---|---|
| With infrastructure shared? → | No | Yes | No | Yes | No | Yes | No | Yes |
| **Capture** | | | | | | | | |
| Total annual capture (MtCO$_2$/y) | 19 | | 23 | | 73 | | 115 | |
| Total number of sources | 53 | | 30 | | 111 | | 194 | |
| Capital investment (billion 2020$) | 12.9 | | 17.5 | | 51.3 | | 81.7 | |
| Average levelized capture cost ($/tCO$_2$) | 99 | | 110 | | 100 | | 102 | |
| **Transport** | | | | | | | | |
| Total pipeline length (kilometers) | 1,709 | 798 | 1,941 | 537 | 6,450 | 1,523 | 10,100 | 2,858 |
| Pipe km per MtCO$_2$/y transported | 90 | 42 | 84 | 23 | 88 | 21 | 88 | 25 |
| Total pipeline capital (billion 2020$) | 1.7 | 1.0 | 2.5 | 0.9 | 9.8 | 3.4 | 14.0 | 5.4 |
| Avg. levelized transport cost ($/tCO$_2$) | 11 | 7 | 14 | 5 | 17 | 6 | 15 | 6 |
| **Storage** | | | | | | | | |
| Storage sites (total number) | 21 | 16 | 10 | 5 | 32 | 24 | 63 | 45 |
| Storage capital (billion 2020$) | 0.8 | 0.7 | 0.7 | 0.7 | 2.3 | 2.4 | 3.8 | 3.8 |
| Average levelized storage cost ($/tCO$_2$) | 11 | 10 | 9 | 8 | 9 | 9 | 9 | 9 |

Close inspection of Figure 11b shows 22 "natural" CCTS hubs, i.e., groups of two or more capture facilities sharing a common T&S network. Only 10 facilities (distributed in the north and southeast), representing 9% of total emissions, are not part of a natural hub because their remoteness makes it sub-optimal from an overall cost perspective to be included in a hub. Table 5 summarizes the characteristics of each of the natural hubs, with additional details provided in SI.B4. Six of the 22 hubs include more than five capture sources, and most of these are in the Southeast region. Only 4 of the hubs, including 3 in the Southeast, require more than one storage site. Thirteen of the hubs each collect and store more than 1 MtCO$_2$/year, and these all have average transport costs less than $1/Mt-km. Hubs that move and store less than 1 MtCO$_2$/y have significantly higher average transport costs.

Smaller facilities (emitting < 1 MtCO$_2$/y), which account for about ¼ of industrial emissions in Louisiana, stand to benefit the most from sharing transportation and storage infrastructure in hubs. To illustrate this, consider hub SE.7, which includes nine capture plants, eight of which capture less than 1 MtCO$_2$/y (see SI.B.SE.7.WO-EJ). The average transport cost across this hub is $6.1/tCO$_2$. For the eight smaller capture plants this represents average savings of more than 85% relative to having dedicated single user pipelines for these facilities. For the largest facility (emitting 2.8 MtCO$_2$/y capture) this would represent a slightly higher cost for transportation than building a dedicated pipeline for its own use (5.7 $/tCO$_2$, see SI.B.3.3. PtP-SE, modeling ID: 68). If the largest facility shared the pipeline system in the hub but was not required to shoulder any of its costs, the average savings in transport costs to the remaining eight facilities would still be nearly 70%.

Table 5. "Natural" hubs, sorted by region and increasing collective per-hub capture amount.

| Region.Hub# (a) | Captured MtCO$_2$/y | # of CO$_2$ sources | # of storage sites | Total pipe-km | Hub-average costs Transport, $/Mt-km | Storage, $/t |
|---|---|---|---|---|---|---|
| North.7 | 0.003 | 2 | 1 | 62 | 40.19 | 11 |
| N.4 | 0.03 | 2 | 1 | 15 | 4.77 | 12 |
| N.9 | 0.04 | 2 | 1 | 37 | 3.75 | 9 |
| N.3 | 0.08 | 4 | 1 | 29 | 1.8 | 12 |
| N.1 | 0.1 | 2 | 1 | 22 | 1.45 | 13 |
| N.5 | 0.1 | 5 | 1 | 68 | 1.27 | 11 |
| N.2 | 1.26 | 4 | 1 | 31 | 0.25 | 12 |
| N.6 | 2.52 | 3 | 1 | 25 | 0.09 | 12 |
| N.8 | 6.13 | 23 | 3 | 402 | 0.03 | 9 |
| Southwest.5 | 0.0002 | 2 | 1 | 207 | 152.71 | 8 |
| SW.1 | 1.47 | 2 | 1 | 45 | 0.12 | 9 |
| SW.3 | 1.83 | 2 | 1 | 27 | 0.1 | 9 |
| SW.2 | 1.94 | 3 | 1 | 110 | 0.08 | 9 |
| SW.4 | 17.41 | 21 | 1 | 148 | 0.03 | 8 |
| Southeast.2 | 0.04 | 4 | 1 | 124 | 3.14 | 8 |
| SE.5 | 0.07 | 3 | 1 | 33 | 3.61 | 9 |
| SE.4 | 1.3 | 5 | 1 | 96 | 0.09 | 8 |
| SE.8 | 2.18 | 2 | 1 | 34 | 0.15 | 8 |
| SE.6 | 2.66 | 2 | 1 | 31 | 0.1 | 8 |
| SE.7 | 4.82 | 9 | 2 | 122 | 0.05 | 8 |
| SE.1 | 12.67 | 17 | 4 | 277 | 0.02 | 9 |
| SE.3 | 47.64 | 65 | 9 | 682 | 0.01 | 9 |

(a) See SI.B4 for location of hubs.

### 4.2.3. Social and environmental justice considerations

The pipeline hubs laid out in Figure 11b were designed without taking into consideration the location of disadvantaged communities, i.e., those identified as marginalized, underserved, or overburdened by pollution. To examine the impact of routing pipelines so as to not reinforce historical practices of infringing on such communities, additional *SimCCS* runs were conducted with the SEJ3 or SEJ8 layer included in the underlying cost surface. This analysis was carried out for Southeast Louisiana, the region of the state with the highest industrial CO$_2$ emissions and population densities. Significantly, the number and makeup of "natural" CCTS hubs changes from the case without SEJ considerations for the same set of capture sites, implying the involvement of different sets of stakeholders in siting/permitting of pipelines (

Figure 12). Total pipeline mileage increases by only 3% (with SEJ8) to 12% (with SEJ3). Large variations in cost-surface weighting factors involved in the routing optimization make it difficult to pinpoint the reason for the smaller increase in pipeline mileage when a larger number of SEJ categories are considered. See SI.B5-B8 for additional details.

The reduced impact on disadvantaged communities is quantified in the table inset in Figure 12, which shows kilometers of pipelines by pipe-flow class that traverse disadvantaged communities when the SEJ layers are activated in the modeling. Total kilometers in disadvantaged areas are reduced by about 80%, with even larger percentage reductions in the largest flow-class pipelines.

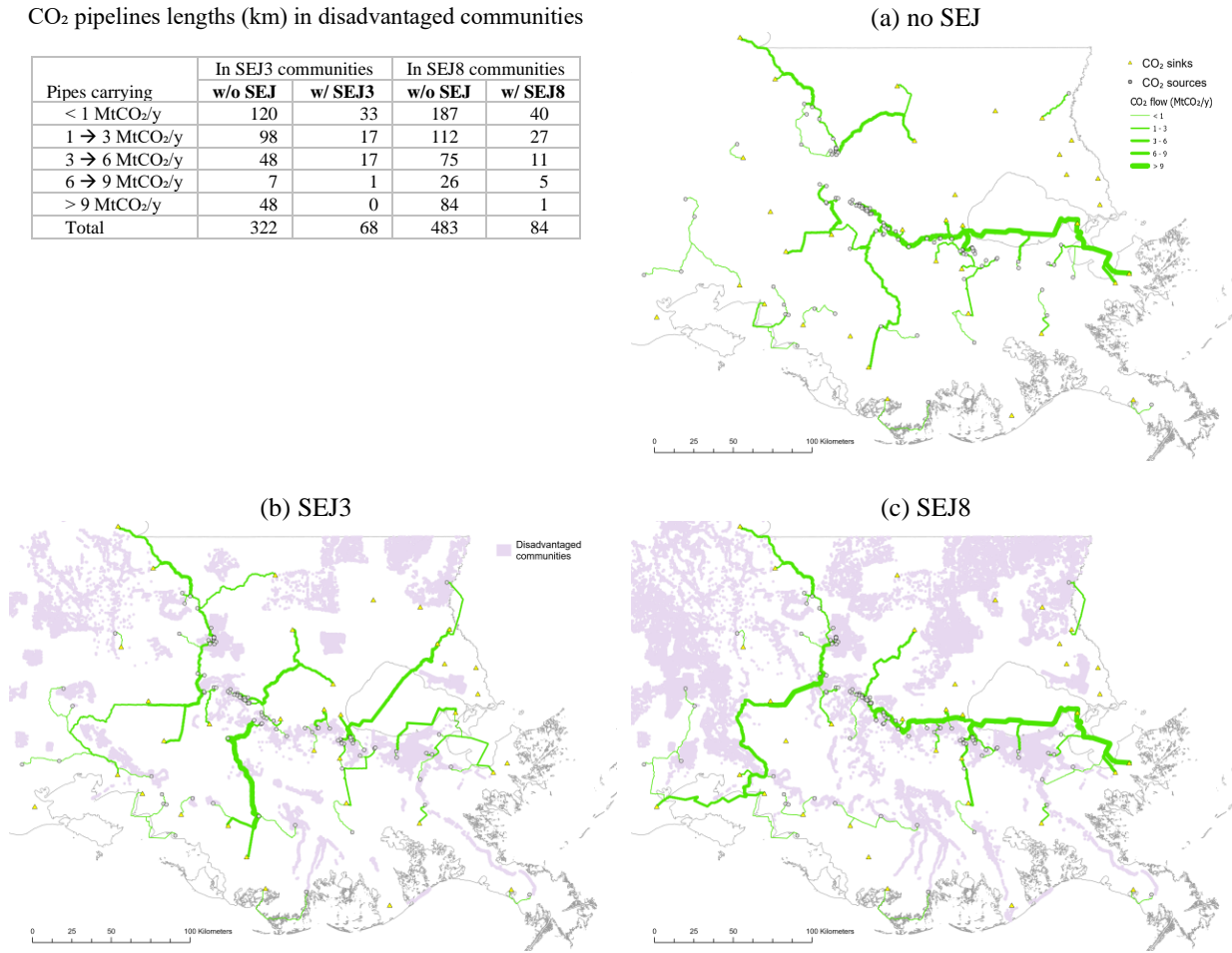

| | In SEJ3 communities | | In SEJ8 communities | |
|---|---|---|---|---|
| Pipes carrying | w/o SEJ | w/ SEJ3 | w/o SEJ | w/ SEJ8 |
| < 1 MtCO$_2$/y | 120 | 33 | 187 | 40 |
| 1 → 3 MtCO$_2$/y | 98 | 17 | 112 | 27 |
| 3 → 6 MtCO$_2$/y | 48 | 17 | 75 | 11 |
| 6 → 9 MtCO$_2$/y | 7 | 1 | 26 | 5 |
| > 9 MtCO$_2$/y | 48 | 0 | 84 | 1 |
| Total | 322 | 68 | 483 | 84 |

CO$_2$ pipelines lengths (km) in disadvantaged communities

Figure 12. Modelled CO$_2$ pipelines for Southeast Louisiana when (a) SEJ is not considered, (b) with SEJ3 layer active, and (c) with SEJ8 layer active. (Circles are CO$_2$ capture points. Triangles are CO$_2$ storage sites.)

### 4.2.4. Building CCTS networks

All the above results describe optimized CO$_2$ capture, transport, and storage networks without regard to the timing of their build out. In this section we describe one approach for analyzing the phasing of network construction. For illustrative purposes, we consider the construction of a network that by 2050 captures and stores CO$_2$ from all current industrial sources emitting more than 100,000 t/y in Southeast Louisiana.[j] In 2050, the system would be capturing and storing 73 Mt/y. We include the SEJ3 layer in pipeline siting considerations, and utilize *SimCCS$^{TIME}$* [66], [67], a variant of *SimCCS$^{PRO}$*, to model construction phases. *SimCCS$^{TIME}$* minimizes the total capital and operating cost for CTS in a region over a specified time duration while meeting exogenous capture/storage targets at intermediate time steps. With perfect foresight, the model may oversize some pipelines in early stages of buildout to facilitate added flow in those pipelines later. (Pipelines can be oversized only up to a specified maximum physical capacity, beyond which a second pipeline following the same routing would be added.) We specify system-wide targets for capture/storage in the Southeast region following a logistics function: 5, 31, 62, 72, and 73 Mt/y in 2030, 2035, 2040, 2045 and 2050, respectively.

---

[j] Industrial emission sources ≤100,000 t/y account for less than 2% of total emissions in Southeast Louisiana.

Figure 13 shows the modeled evolution of the CCTS system. Each map shows in-service plant and infrastructure at five-year timesteps, starting in 2030. In the first phase, a single ammonia-producing facility – the largest current emitter in Southeast Louisiana – is capturing 5 Mt/y of its process $CO_2$ and sending it to two different storage sites via a total of 84 km of pipelines. By 2035 an additional 11 capture plants are capturing 24 $MtCO_2$/y, and 11 additional storage sites are active. A very rapid expansion then occurs to 2040, when 24 additional capture plants and 6 additional storage sites are in service, along with a cumulative 929 km of pipelines. The expansion begins to slow thereafter, with 20 more capture plants and 4 more storage sites operating by 2045. Finally, by 2050, 12 more capture plants and one additional storage site are online.

The cumulative "overnight" capital required for the full buildout is about $50 billion for capture plants and about $6 billion for transportation and storage infrastructure (in 2020$), with disbursements of capital as shown in Figure 14a. Factoring in four-year construction times and 30-year operating lifetimes, Figure 14b shows the average levelized cost for capture, transport, and storage at each phase of the buildout. In phase 1, the high concentration and large capture volume of $CO_2$ for the single capture plant operating that year results in a levelized cost of $39/$tCO_2$ before factoring in the Inflation Reduction Act (IRA) 45Q tax credit. With the 45Q tax credit of $85/t available for the first 12 years of operation[k] [68] the levelized cost is negative. The average cost without the tax credit for Phase 2 exceeds $134/$tCO_2$. Phase 2 projects are assumed to start construction prior to 2033 and so quality for the 45Q tax credit, which reduces the average levelized cost for projects coming online in 2035 to $73/$tCO_2$. The IRA tax credit

---

[k] We assume that all facilities pay prevailing wages during the construction period and first 12 years of operation and meet registered apprenticeship requirements, thus qualifying for the $85/$tCO_2$ level of 45Q credit.

for projects that begin construction after 2032 is $17/t, and so the IRA credit impacts the levelized cost in the final three phases only modestly: net costs range from $145 to $207/tCO$_2$.

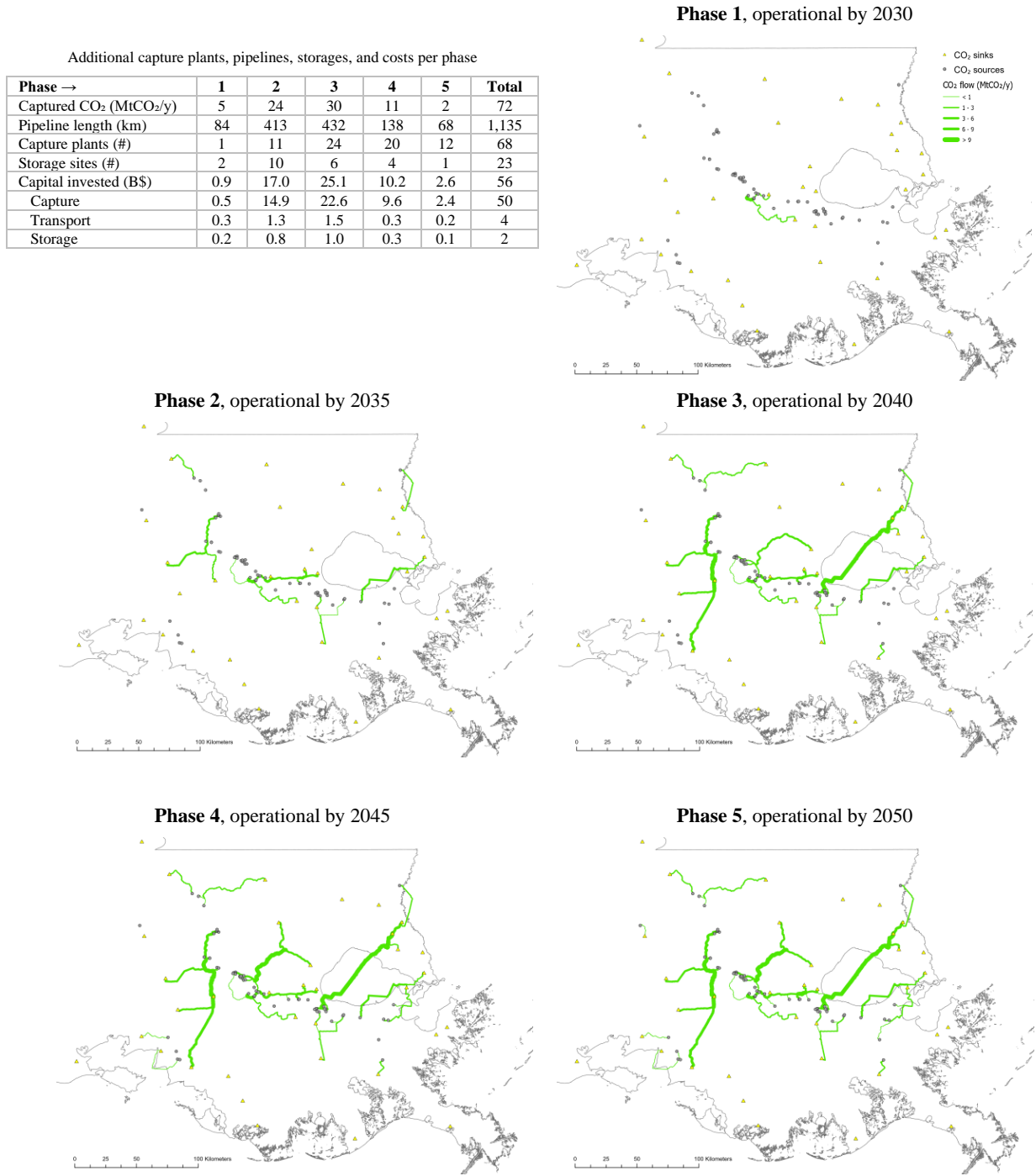

| Additional capture plants, pipelines, storages, and costs per phase | | | | | | |
|---|---|---|---|---|---|---|
| Phase → | 1 | 2 | 3 | 4 | 5 | Total |
| Captured CO$_2$ (MtCO$_2$/y) | 5 | 24 | 30 | 11 | 2 | 72 |
| Pipeline length (km) | 84 | 413 | 432 | 138 | 68 | 1,135 |
| Capture plants (#) | 1 | 11 | 24 | 20 | 12 | 68 |
| Storage sites (#) | 2 | 10 | 6 | 4 | 1 | 23 |
| Capital invested (B$) | 0.9 | 17.0 | 25.1 | 10.2 | 2.6 | 56 |
| Capture | 0.5 | 14.9 | 22.6 | 9.6 | 2.4 | 50 |
| Transport | 0.3 | 1.3 | 1.5 | 0.3 | 0.2 | 4 |
| Storage | 0.2 | 0.8 | 1.0 | 0.3 | 0.1 | 2 |

Figure 13. Cost-minimized expansion of a CCTS system for Southeast Louisiana. Operating plant and infrastructure are shown in five-year time steps from 2030 (phase 1) to 2050 (phase 5).

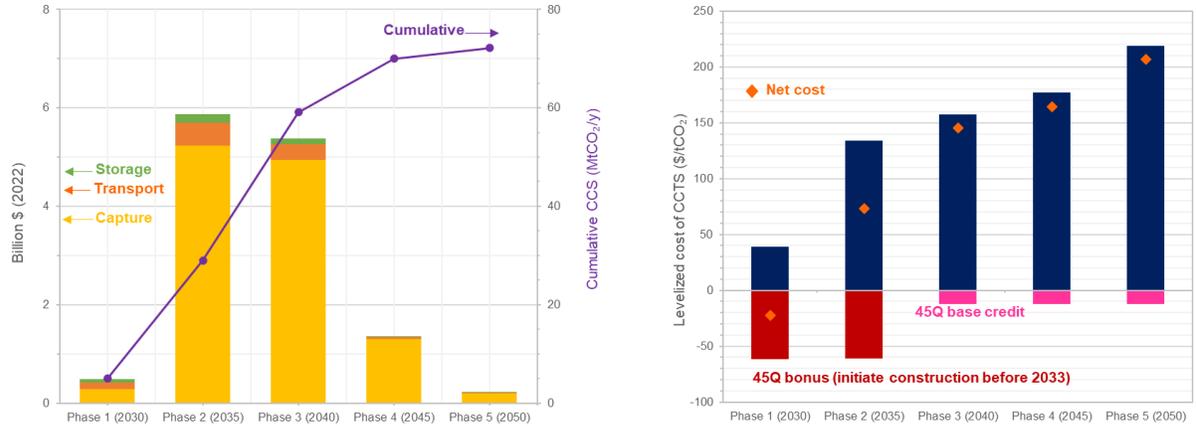

Figure 14. (a) Capital investment required and (b) average levelized cost per tCO₂ captured and stored for each phase of the CCTS network buildout depicted in Figure 13.

## 5. Summary and future directions

We developed methodologies for, and evaluated, the design of CO₂ capture, transport, and storage (CCTS) systems serving existing industrial CO₂ emitters, with a case study focus on Louisiana. The work included evaluating the scale and concentration of capturable CO₂ emissions at individual industrial facilities to enable estimating capital and operating costs for CO₂ capture retrofits. For CO₂ sinks, a screening method was developed to identify potential CO₂ storage sites considering both above-ground and subsurface features and then estimating site storage capacities, injectivities, and costs. Cost-minimized pipeline network designs subject to land use patterns, existing rights-of-way, demographics, and other social and environmental constraints, were developed to connect CO₂ capture plants with storage sites identified sinks are connected via cost-minimized CO₂ pipeline networks.

We identified 130 million tonnes per year (Mt/y) of existing CO₂ emissions from 190 facilities in Louisiana, nearly 80% of which would be capturable for $120/tCO₂ or less. A total of 98 potential storage sites were identified, with aggregate injectivity between 327 and 731 MtCO₂/y and storage costs ranging from $8 to $17/tCO₂. We compared the features of optimal (cost-minimized) pipelines that deliver CO₂ from individual capture plants to storage sites against the features of pipeline networks shared by multiple capture plants. When considering all industrial facilities in the state, the shared networks reduced total aggregate length of pipelines by more than 75%, average unit transport cost by more than 60%, and the number of storage sites required by nearly 30%. Smaller facilities (capturing < 1 MtCO₂/y), which account for about ¼ of Louisiana's industrial emissions today, stand to benefit the most from sharing of transportation and storage

networks. Networked infrastructure designs that explicitly avoid routing pipelines through disadvantaged (environmental justice) communities, so as to not reinforce historical damages, involve only modestly higher pipeline lengths and costs than networks that disregard this consideration.

The cumulative "overnight" capital needed for constructing an optimal CCTS system serving all industrial facilities emitting more than 100,000 t/y each in Southeast Louisiana (and capturing 73 Mt$CO_2$/y) is estimated to be $56 billion (2020$), of which 90% is for the capture plants. Phasing-in this investment in five-year time steps from 2030 to 2050 and including the bonus 45Q tax credit ($85/t$CO_2$) in the initial two time steps could result in net negative levelized costs per t$CO_2$ initially, but higher costs in the future, as higher capture-cost facilities are added and the 45Q bonus incentive reverts to the 45Q base level ($17/t$CO_2$).

Our study suggests several directions for future investigations:

- Our analysis did not consider $CO_2$ storage offshore (sub-seabed). Pipeline construction costs may be higher than onshore in some cases, but construction approvals may be easier to obtain. Trade-offs with on-shore versus off-shore storage would be of interest to investigate.

- We noted earlier that $CO_2$ capture retrofits at industrial facilities likely will bring health benefits to people in nearby surrounding areas due to concomitant reductions in emissions of local pollutants. Quantifying and overlaying such public health benefits on the cost analysis of CCTS hubs here would further help inform public policies aimed at optimizing the deployment of CCTS.

- Our analyses assumed on-site natural gas cogeneration would provide parasitic heat and electricity needs for $CO_2$ capture plants. Ongoing grid decarbonization and use of low-carbon cogeneration fuels, e.g., hydrogen, could reduce or eliminate the attendant "carbon penalty". Investigation and evaluation of such options would be of interest.

- Our analysis here focused on existing industrial facilities in Louisiana without consideration of any future growth. Growth scenarios would be of interest to consider, along with how CCTS systems might most optimally be integrated with other industrial decarbonization strategies, including process efficiency improvements and electrification, zero- or low-emission fuels and feedstocks, and (for negative emissions) direct air capture of $CO_2$ and bioenergy with CCS.

## 6. Conflicts of interest

There are no conflicts of interest to declare.

## 7. Acknowledgements

The authors thank the following individuals for helpful consultations in the course of our work: Richard Middleton, Erin Middleton, Carl Tasma and Jonathan Ogland-Hand (Carbon Solutions LLC), Tim Barckholtz (ExxonMobil), and George Walchuk (ExxonMobil, retired). The authors gratefully acknowledge financial support for this work from Deloitte, Weyerhaeuser, ExxonMobil, Princeton University's Andlinger Center for Energy and the Environment, and Princeton's (BP-funded) Carbon Mitigation Initiative.